# Broadband Polarizers Based on Graphene Metasurfaces

TIANJING GUO AND CHRISTOS ARGYROPOULOS*

*Department of Electrical and Computer Engineering, University of Nebraska-Lincoln, Lincoln, NE, 68588, USA*
*Corresponding author: christos.argyropoulos@unl.edu*

**We present terahertz (THz) metasurfaces based on aligned rectangular graphene patches placed on top of a dielectric layer to convert the transmitted linearly polarized waves to circular or elliptical polarized radiation. Our results lead to the design of an ultrathin broadband THz quarter-wave plate. In addition, ultrathin metasurfaces based on arrays of L-shaped graphene periodic patches are demonstrated to achieve broadband cross-polarization transformation in reflection and transmission. The proposed metasurface designs have tunable responses and are envisioned to become the building blocks of several integrated THz systems.**

Controlling the polarization state of electromagnetic waves is of high practical importance to THz communications, imaging and sensing [1]. It holds great promise to the advancement of many fields, such as optoelectronics, analytical chemistry, and biology. In recent years, metamaterials have opened up an exciting route for manipulating the polarization of electromagnetic waves [2] because of their ultrathin profiles and efficient operation compared to naturally available polarization crystals. In particular, chiral metamaterials (CMMs) [3] have been widely used to achieve polarization conversion. It has been demonstrated that bilayer CMMs or multilayer CMMs can accomplish high-efficiency broadband polarization control [4, 5]. However, CMMs usually require complicated fabrication processes to be constructed due to their thin thickness and need of precise alignment between their different elements. In addition, it is difficult to have a dynamic reconfigurable operation and suffer from increased Ohmic losses.

Metasurfaces [6], the planar counterparts of metamaterials, are destined to have more widespread applications because of several intriguing properties, such as lower loss, compact size, and higher efficiency. They can provide a large degree of control over the amplitude, phase, and polarization of electromagnetic waves. For example, by splitting the incident waves into two orthogonal components and obtaining the required phase delay between them, metasurfaces can be utilized to manipulate the polarization of electromagnetic waves. To this end, plasmonic metasurfaces have been demonstrated to realize ultrathin quarter-wave plates [7]. However, these devices cannot be made tunable and broadband and their fabrication is still complicated.

Alternative metasurfaces based on different materials, such as graphene, can be used to design easier to implement and broadband ultrathin polarizers. Graphene is a two-dimensional (2D) layer of carbon atoms arranged in a honeycomb lattice. Its outstanding properties made it an excellent material for future electronic and photonic devices. Various graphene-based devices have recently been demonstrated: infrared imaging sensors [8], cloaks [9], and modulators [10-12]. Graphene's extraordinary electronic and optical properties [13] can lead to novel ways to enhance and control the interaction between light and matter. Recently, graphene metasurfaces have been created by patterning a graphene monolayer placed on top of an insulator [14, 15]. These new type of metasurfaces usually operate at low THz frequencies, where graphene exhibits strong plasmonic behavior.

In this Letter, we present a new design of a broadband ultrathin linear-to-circular or elliptical polarization converter (quarter-wave plate) operating in transmission based on periodic rectangular graphene patches placed over a dielectric substrate. In addition, we propose a new broadband ultrathin cross-polarization converter, i.e. half-wave plate, operating in reflection composed of an array of L-shaped graphene patches placed over a dielectric spacer layer backed by a gold substrate to suppress the transmission. The former can convert linear to circular or elliptical polarized waves in transmission and the latter can convert linear polarized waves to their cross polarized version in reflection. Compared to previous works, the proposed THz polarizers are more simple to experimentally verify, they have an ultrathin thickness, and can work over a wide frequency range and for different angles of the incident wave. In addition, their properties can be dynamically tuned along the entire THz frequency spectrum. The intraband and interband transitions contribute to the conductivity of graphene in different ways. In the low THz frequency region, we can ignore the interband transitions and the complex intraband-based conductivity is modeled by the Drude expression [16], $\sigma = -je^2 E_F / (\pi \hbar^2 (\omega - j2\Gamma))$, where $\Gamma$ is the electron-phonon scattering rate due to the carrier intraband scattering described as $\Gamma = ev_F^2 / \mu E_F$, where $v_F \approx c/300$ is the Fermi velocity, $E_F$ is the Fermi level or doping level, $\mu = 1 m^2/Vs$ is the measured DC mobility [17] and the relaxation time $\tau = 1/\Gamma$ is approximately $\tau = 1 ps$. The permittivity of graphene can be expressed as $\varepsilon = 1 + i\sigma / (\varepsilon_0 \omega \delta)$, where $\delta = 1 nm$ is the typical thickness of graphene used in simulations. In addition, the permittivity of gold follows a Drude model expression fitting the experimental data [18], $\varepsilon_{Au} = \varepsilon_\infty - f_p^2 / [f(f + i\gamma)]$ with plasma frequency $f_p = 2069 THz$, damping constant $\gamma = 17.65 THz$ and $\varepsilon_\infty = 1.53$. The permittivity of the dielectric substrate layer is assumed to be equal to 2.25, similar to silica.

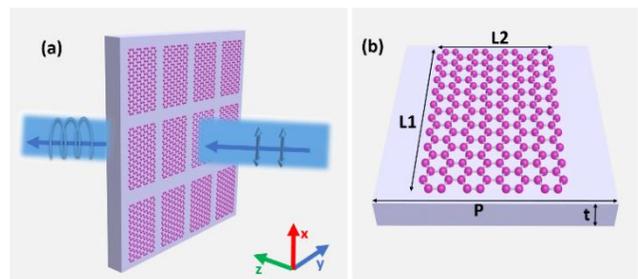

Fig. 1. (a) Schematic of an ultrathin quarter-wave plate metasurface made of an array of rectangular graphene patches. (b) Square unit cell of the design with dimensions: P=7.6um, L1=6.5um, L2=5.2um, t=1.5um.

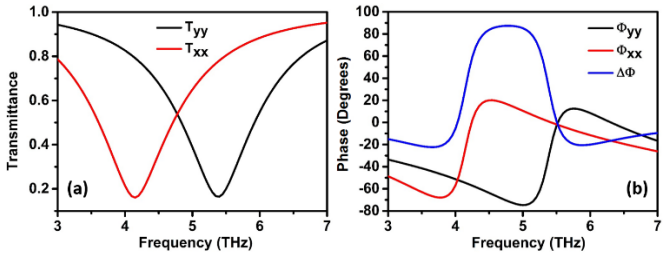

Fig. 2. (a) Transmission coefficients and (b) phase difference for the x- and y-polarized normally incident waves.

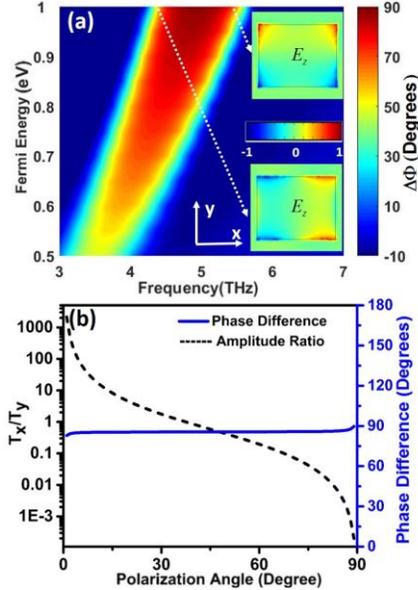

Fig. 3. (a) The calculated phase difference of the circular polarization converter as a function of operation frequency and Fermi energy. Insets: the z component of the electric field distribution at the two resonances. (b) Transmission amplitude ratio (black) and phase difference (blue) at the 4.75THz resonance frequency of the proposed quarter-wave plate.

The schematic depiction of the proposed ultrathin quarter-wave plate is shown in Fig. 1 with dimensions given in the caption. P represents the periodicity, L1 and L2 represent the length and width of the graphene patches, respectively, and t is the thickness of the dielectric substrate. These geometrical parameters have been optimized using simulations based on COMSOL Multiphysics. In our simulations, the excitation source is considered to be a linearly polarized plane wave with either x or y polarization. The Fermi energy of graphene is equal to 0.95eV. The transmission coefficients are defined as $T_{ij} = |E_j^{Trans} / E_i^{Inc}| (i, j = x, y)$, where $E_j^{Inc} (i = x, y)$ is the x- or y-polarized incident waves and $E_j^{Trans} (j = x, y)$ is the x or y component of the transmitted waves, respectively [19]. The phase is defined as $\Phi_{ij} = \arg(E_j^{Trans} / E_i^{Inc})(i, j = x, y)$. The transmission coefficients and phases of each polarization are reported in Fig. 2. The phase difference $\Delta\Phi = \Phi_{xx} - \Phi_{yy}$ is also plotted with the blue line in Fig. 2(b). We emphasize that the proposed metasurface has negligible cross-polarization conversion.

The transmission resonance is shifted when the proposed ultrathin linear-to-circular or elliptical transformer is illuminated by different orthogonal linear polarizations aligned either along the x or y directions. These two distinct resonances arise from the asymmetry between the length and width of the graphene patches. The bigger is the difference between L1 and L2, the larger frequency shift will be obtained for different orthogonal linear polarizations. The transmission coefficients of both orthogonal linear polarized incident waves are the same at frequency 4.75THz while the phase difference is exactly 90°, leading to the realization of an ultrathin quarter-wave plate. Interestingly, the 90° phase difference can be obtained in a broad frequency range, from 4.5THz to 5.3THz, as shown in Fig. 2(b), with the drawback of reduced transmission.

The phase difference of the proposed ultrathin polarization converter can be tuned by controlling the Fermi level (doping) of graphene, as reported in Fig. 3(a). The proposed quarter-wave plate is working in a broad frequency range and its operation blueshifts as the Fermi level increases. In general, the resonant frequency of periodic rectangular graphene patches has been demonstrated to be affected by their length $L$ and the Fermi level $E_F$ following the relationship $f_r \propto \sqrt{E_F / L}$ [20], consistent with the blueshift obtained in Fig. 3(a). The z component of the electric field distributions at the two resonances are calculated and shown in the inset of Fig. 3(a). Dipolar resonant modes are excited along the edges of the graphene patches with different orientations depending on the polarization of the incident wave. Finally, the incident polarization angle is swept from 1° to 89° at the resonance frequency (4.75THz). The amplitude ratio and phase difference between the two orthogonal linear polarizations is calculated to verify that the proposed circular or elliptical polarization converter is not sensitive to the incident wave angles. The results are shown in Fig. 3(b) and we can conclude that elliptical polarized transmitted light can be obtained for all incident angles, since always $\Delta\Phi = 90°$ and $T_{xx}/T_{yy} \neq 1$, and circular polarized transmitted light can be obtained when the incident polarization angle is 45° and $T_{xx}/T_{yy} = 1$. Note that the proposed metasurface can also operate inversely, i.e. circular or elliptical polarized incident light will be transformed to linear polarized light with different polarization angles.

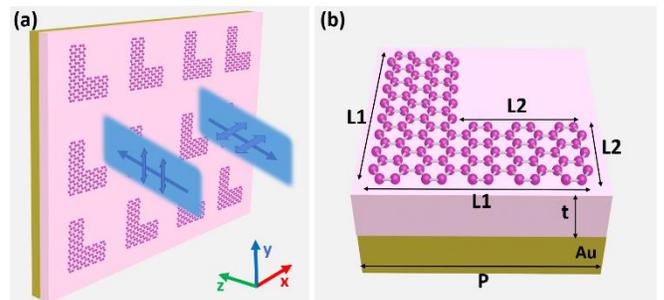

Fig. 4. (a) Schematic of an ultrathin half-wave plate made of an array of L-shaped graphene patches. (b) Square unit cell of the proposed design with dimensions: P=3.6um, L1=2.4um, L2=1.2um, t=7.5um.

Next, we propose an ultrathin half-wave plate based on a graphene metasurface with schematic shown in Fig. 4 and parameters reported in the caption. In this case, the metasurface is terminated by a gold substrate and the device operates in reflection. The intermediate dielectric layer has the same permittivity with the previous metasurface. The reflection coefficient can be defined as $R_{ij} = |E_j^{Reflec}/E_i^{Inc}|$ $(i,j = x,y)$ where $E_i^{Inc}$ $(i = x,y)$ is the electric field of the x- or y-polarized incident waves and $E_j^{Reflec}$ $(j = x,y)$ is the x or y component of the reflected waves, respectively. Using full-wave simulations, we calculate the co- and cross-polarization reflection coefficients when an x-polarized wave is incident ($R_{xx}$, $R_{xy}$). The computed $R_{xx}$ and $R_{xy}$ are shown in Fig. 5(a), where the Fermi level of graphene is 0.9eV. In Fig. 5(a), we observe that the cross-polarization $R_{xy}$ is much larger than the co-polarization $R_{xx}$ in a broad frequency range due to critical coupling between the two orthogonally polarized incident waves. Hence, we define the polarization conversion rate (PCR) as $PCR = |R_{xy}|^2/[|R_{xx}|^2 + |R_{xy}|^2]$ [21] to reveal the efficiency of the proposed cross-polarization converter (half-wave plate). The phase difference is also defined as $\Delta\Phi = \Phi_{xy} - \Phi_{xx}$. The calculated PCR and phase difference of the L-shaped graphene metasurface is plotted in Fig. 5(b). It can be seen that the PCR is always above 70% from 5.09THz to 8.25THz and approaches almost 1.0 at the two resonance frequencies of 5.4THz and 8THz. At 5.4THz: PCR= 0.975, $\Delta\Phi = 0°$, and at 8THz: PCR=0.966, $\Delta\Phi = 180°$, which leads to a dual-band half-wave plate operation in these resonance frequencies. The changes in phase difference at these particular two resonance frequencies lead to even and odd resonant modes along the graphene surface due to the strong cross-polarization coupling effect. The field profiles of these different resonant surface modes can be seen by their normalized electric field distributions, shown in the inset of Fig. 6(a).

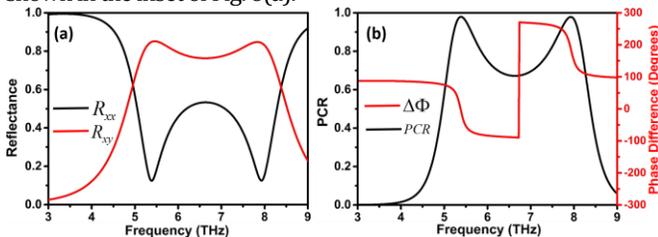

Fig. 5. (a) Co- (black) and cross- (red) polarized reflection coefficients. (b) PCR and phase difference for the ultrathin half-wave plate.

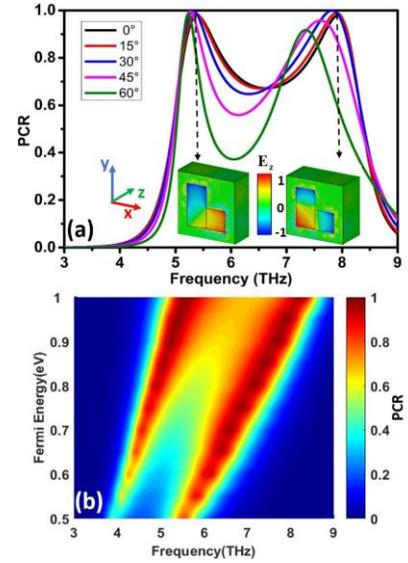

Fig. 6. (a) Calculated PCR under different incident angle illuminations. Insets: the z component of the normalized electric field distribution at the first resonance (even mode) and the second resonance (odd mode). (b) The calculated PCR of the cross-polarization graphene metasurface converter as a function of the operation frequency and Fermi energy.

The performance of the proposed cross-polarizer is angle insensitivity, as verified in Fig. 6(a). This is an important feature for polarization converters. We calculate the PCRs for this L-shaped polarization converter under oblique incident angle illuminations ranging from 0° to 60°. The PCR at the first resonance frequency is almost invariable to the incident angle, while the PCR at the second resonance frequency slightly declines as the incident angles increase to values larger than 30°. The even mode has a symmetric field distribution and, as a result, is relative insensitive to the incident angles. The odd mode has an asymmetric field profile, which makes it more sensitive to the angles of the incident wave. The PCR has values above 0.65 in a wide frequency range for incident angles less than 30°, which indicates a broadband, angle-insensitive polarization converter. Note that the effect of oblique incident waves in phase difference is negligible and we always obtain phase differences of 0° and 180° at two different resonance frequencies, similar to Fig. 5(b). The z components of the normalized electric field distributions at the two resonances are also shown in the inset of Fig. 6(a). The left field distribution corresponds to the first resonant frequency (5.4THz) and the right distribution to the second resonance (8THz).

We also calculate the PCR for different graphene Fermi energies in order to demonstrate the tunability of the proposed ultrathin cross-polarization converter. The results are shown in Fig. 6(b). The proposed converter works at a broad frequency range and the PCR increases as the Fermi energy is increased. This is due to graphene's more plasmonic (metallic) behavior under higher doping levels [22]. Moreover, Fig. 6(b) demonstrates that both resonant frequencies blue shift as the Fermi energy is increased, similar to the results in Fig. 3(a). The quality factor of the resonances appears to deteriorate at higher Fermi levels because of the competition from the narrowing in graphene's conductivity due to the increased Fermi level and the broadening in the loss function coming from the blue shift of the resonances. In addition, the ellipticity angle varies between -10° and 10° at the frequency range of interest and its spectrum (not shown here) also blue shifts as the Fermi level is increased. The tunability of the proposed graphene-based devices is a major advantage compared to metal-based devices.

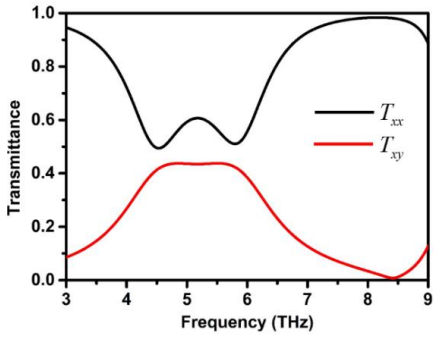

Fig. 7. Co- and cross-polarized transmission coefficients $T_{xx}$ and $T_{xy}$ of the L-shaped graphene polarization converter shown in Fig. 4 after removing the bottom gold layer.

Furthermore, we remove the bottom gold substrate and analyze the proposed cross-polarization device shown in Fig. 4, now, operating in transmission. Applying reciprocity and electromagnetic boundary conditions in this non-magnetic, passive, and ultrathin subwavelength structure, the maximum value of the amplitude of the cross-polarization transmission coefficient can be determined by the equation: $|T_{xx}|^2 - \mathrm{Re}[T_{xx}] + |T_{xy}|^2 = 0$ [23], where $T_{xx}$ is the transmission of the co-polarized wave and $T_{xy}$ is the transmission of the cross-polarized wave. This equation leads to the theoretical maximum limit of cross-polarization transmission for this ultrathin graphene metasurface. The maximum attainable cross-polarization value is $|T_{xy}| = 0.5$ when $|T_{xx}| = 0.5$. The computed maximum cross-polarization transmission of the proposed graphene metasurface is $|T_{xy}| = 0.43$, as shown in Fig. 7, which is slightly lower to the upper theoretical limit due to the Ohmic losses of graphene.

Finally, we provide an insight on the voltage values needed to obtain the currently used doping levels in a potential experimental verification of the proposed graphene metasurfaces. The Fermi energy used in our previous examples can be achieved by electrostatically gating the graphene monolayer with transparent electrodes placed between the graphene and the dielectric layer. Hence, the applied gate voltages can alter the resistivity and conductivity of graphene. In the case of moderately doped graphene, i.e. $E_F \gg k_B T$, the Fermi energy can be obtained by the formula $E_F = \hbar v_F \sqrt{\pi C V_g / e}$ [17], where $k_B$ is the Boltzmann constant, $T$ is the temperature, $\hbar$ is the Planck constant and $V_g$ is the gate voltage between the electrodes. The electrostatic capacitance per unit area can be described as $C = \varepsilon_d \varepsilon_0 / t$, where $\varepsilon_d$ and $t$ is the static permittivity and thickness of dielectric layer, respectively. For the linear-to-circular or elliptical polarization converter with thickness 1500nm, the necessary voltage is approximately 114V to achieve Fermi energy equal to 0.95eV. This voltage value is relative low, will not result in dielectric breakdown, and has been used in previous graphene experiments [24].

In conclusion, ultrathin broadband linear-to-circular or elliptical polarizers and cross-polarization converters have been designed operating at THz frequencies. Compared to previous works, the proposed THz polarizers are more simple to experimentally verify, they have an ultrathin thickness, and can work over a wide frequency range and for different angles of the incident waves. In addition, their properties can be dynamically tuned along the entire THz frequency spectrum without changing their geometry and just by electrostatically gating graphene. The presented results can lead to new tunable electro-optical devices operating at low THz frequencies. Owing to their simple, compact, and tunable design, the proposed graphene-based broadband polarizers are envisioned to become the building blocks of future integrated THz systems.

## Acknowledgments
This work was partially supported by the Office of Research and Economic Development at University of Nebraska-Lincoln, Nebraska EPSCoR, and NSF Nebraska MRSEC.

## References
1. M. Born and E. Wolf, Principles of Optics, 7th ed. (Cambridge University, 1999).
2. N. K. Grady, J. E. Heyes, D. R. Chowdhury, Y. Zeng, M. T. Reiten, A. K. Azad, A. J. Taylor, D. a R. Dalvit, and H.-T. Chen, Science **340**, 1304 (2013).
3. B. Wang, J. Zhou, T. Koschny, M. Kafesaki, and C. M. Soukoulis, J. Opt. A Pure Appl. Opt. **11**, 114003 (2009).
4. K. Song, Y. Liu, C. Luo, and X. Zhao, J. Phys. D. Appl. Phys. **47**, 505104 (2014).
5. Y. Ye and S. He, Appl. Phys. Lett. **96**, 203501 (2010).
6. C. L. Holloway, E. F. Kuester, J. A. Gordon, J. O'Hara, J. Booth, and D. R. Smith, IEEE Antennas Propag. Mag. **54**, 10 (2012).
7. Y. Zhao and A. Alù, Phys. Rev. B **84**, 205428 (2011).
8. X. Sun, Z. Liu, K. Welsher, J. T. Robinson, A. Goodwin, S. Zaric, and H. Dai, Nano Res. **1**, 203 (2008).
9. P. Chen and A. Alù, ACS Nano **5**, 5855 (2011).
10. C. Argyropoulos, Opt. Express **23**, 23787 (2015).
11. M. Liu, X. Yin, E. Ulin-Avila, B. Geng, T. Zentgraf, L. Ju, F. Wang, and X. Zhang, Nature **474**, 64 (2011).
12. Z. Miao, Q. Wu, X. Li, Q. He, K. Ding, Z. An, Y. Zhang, and L. Zhou, Phys. Rev. X **5**, 41027 (2015).
13. K. S. Novoselov, A. K. Geim, S. V Morozov, D. Jiang, M. I. Katsnelson, I. V Grigorieva, S. V Dubonos, and A. A. Firsov, Nature **438**, 197 (2005).
14. H. Yan, X. Li, B. Chandra, G. Tulevski, Y. Wu, M. Freitag, W. Zhu, P. Avouris, and F. Xia, Nat. Nanotechnol. **7**, 330 (2012).
15. L. Ju, B. Geng, J. Horng, C. Girit, M. Martin, Z. Hao, H. A. Bechtel, X. Liang, A. Zettl, Y. R. Shen, and F. Wang, Nat. Nanotechnol. **6**, 630 (2011).
16. P.-Y. Chen, C. Argyropoulos, and A. Alu, IEEE Trans. Antennas Propag. **61**, 1528 (2013).
17. K. S. Novoselov, Science **306**, 666 (2004).
18. P. B. Johnson and R. W. Christy, Phys. Rev. B **6**, 4370 (1972).
19. H. Cheng, S. Chen, P. Yu, J. Li, L. Deng, and J. Tian, Opt. Lett. **38**, 1567 (2013).
20. J. Ding, B. Arigong, H. Ren, J. Shao, M. Zhou, Y. Lin, and H. Zhang, Plasmonics **10**, 351 (2015).
21. J. Hao, Q. Ren, Z. An, X. Huang, Z. Chen, M. Qiu, and L. Zhou, Phys. Rev. A **80**, 23807 (2009).
22. C. Yang, Y. Luo, J. Guo, Y. Pu, D. He, Y. Jiang, J. Xu, and Z. Liu, Opt. Express **24**, 16913 (2016).
23. F. Monticone, N. M. Estakhri, and A. Alù, Phys. Rev. Lett. **110**, 203903 (2013).
24. S. H. Lee, M. Choi, T.-T. Kim, S. Lee, M. Liu, X. Yin, H. K. Choi, S. S. Lee, C.-G. Choi, S.-Y. Choi, X. Zhang, and B. Min, Nat. Mater. **11**, 936 (2012).

## References

1. M. Born and E. Wolf, Principles of Optics, 7th ed. (Cambridge University, 1999).
2. N. K. Grady, J. E. Heyes, D. R. Chowdhury, Y. Zeng, M. T. Reiten, A. K. Azad, A. J. Taylor, D. a R. Dalvit, and H.-T. Chen, "Terahertz Metamaterials for Linear Polarization Conversion and Anomalous Refraction," Science **340**, 1304–1307 (2013).
3. B. Wang, J. Zhou, T. Koschny, M. Kafesaki, and C. M. Soukoulis, "Chiral metamaterials: simulations and experiments," J. Opt. A Pure Appl. Opt. **11**, 114003 (2009).
4. K. Song, Y. Liu, C. Luo, and X. Zhao, "High-efficiency broadband and multiband cross-polarization conversion using chiral metamaterial," J. Phys. D. Appl. Phys. **47**, 505104 (2014).
5. Y. Ye and S. He, "90° polarization rotator using a bilayered chiral metamaterial with giant optical activity," Appl. Phys. Lett. **96**, 203501 (2010).
6. C. L. Holloway, E. F. Kuester, J. A. Gordon, J. O'Hara, J. Booth, and D. R. Smith, "An Overview of the Theory and Applications of Metasurfaces: The Two-Dimensional Equivalents of Metamaterials," IEEE Antennas Propag. Mag. **54**, 10–35 (2012).
7. Y. Zhao and A. Alù, "Manipulating light polarization with ultrathin plasmonic metasurfaces," Phys. Rev. B **84**, 205428 (2011).
8. X. Sun, Z. Liu, K. Welsher, J. T. Robinson, A. Goodwin, S. Zaric, and H. Dai, "Nano-graphene oxide for cellular imaging and drug delivery," Nano Res. **1**, 203–212 (2008).
9. P. Chen and A. Alù, "Atomically Thin Surface Cloak Using Graphene Monolayers," ACS Nano **5**, 5855–5863 (2011).
10. C. Argyropoulos, "Enhanced transmission modulation based on dielectric metasurfaces loaded with graphene," Opt. Express **23**, 23787 (2015).
11. M. Liu, X. Yin, E. Ulin-Avila, B. Geng, T. Zentgraf, L. Ju, F. Wang, and X. Zhang, "A graphene-based broadband optical modulator," Nature **474**, 64–67 (2011).
12. Z. Miao, Q. Wu, X. Li, Q. He, K. Ding, Z. An, Y. Zhang, and L. Zhou, "Widely Tunable Terahertz Phase Modulation with Gate-Controlled Graphene Metasurfaces," Phys. Rev. X **5**, 41027 (2015).
13. K. S. Novoselov, A. K. Geim, S. V Morozov, D. Jiang, M. I. Katsnelson, I. V Grigorieva, S. V Dubonos, and A. A. Firsov, "Two-dimensional gas of massless Dirac fermions in graphene," Nature **438**, 197–200 (2005).
14. H. Yan, X. Li, B. Chandra, G. Tulevski, Y. Wu, M. Freitag, W. Zhu, P. Avouris, and F. Xia, "Tunable infrared plasmonic devices using graphene/insulator stacks," Nat. Nanotechnol. **7**, 330–334 (2012).
15. L. Ju, B. Geng, J. Horng, C. Girit, M. Martin, Z. Hao, H. A. Bechtel, X. Liang, A. Zettl, Y. R. Shen, and F. Wang, "Graphene plasmonics for tunable terahertz metamaterials," Nat. Nanotechnol. **6**, 630–634 (2011).
16. P.-Y. Chen, C. Argyropoulos, and A. Alu, "Terahertz Antenna Phase Shifters Using Integrally-Gated Graphene Transmission-Lines," IEEE Trans. Antennas Propag. **61**, 1528–1537 (2013).
17. K. S. Novoselov, "Electric Field Effect in Atomically Thin Carbon Films," Science **306**, 666–669 (2004).
18. P. B. Johnson and R. W. Christy, "Optical Constants of the Noble Metals," Phys. Rev. B **6**, 4370–4379 (1972).
19. H. Cheng, S. Chen, P. Yu, J. Li, L. Deng, and J. Tian, "Mid-infrared tunable optical polarization converter composed of asymmetric graphene nanocrosses.," Opt. Lett. **38**, 1567–1569 (2013).
20. J. Ding, B. Arigong, H. Ren, J. Shao, M. Zhou, Y. Lin, and H. Zhang, "Mid-Infrared Tunable Dual-Frequency Cross Polarization Converters Using Graphene-Based L-Shaped Nanoslot Array," Plasmonics **10**, 351–356 (2015).
21. J. Hao, Q. Ren, Z. An, X. Huang, Z. Chen, M. Qiu, and L. Zhou, "Optical metamaterial for polarization control," Phys. Rev. A **80**, 23807 (2009).
22. C. Yang, Y. Luo, J. Guo, Y. Pu, D. He, Y. Jiang, J. Xu, and Z. Liu, "Wideband tunable mid-infrared cross polarization converter using rectangle-shape perforated graphene," Opt. Express **24**, 16913 (2016).
23. F. Monticone, N. M. Estakhri, and A. Alù, "Full Control of Nanoscale Optical Transmission with a Composite Metascreen," Phys. Rev. Lett. **110**, 203903 (2013).
24. S. H. Lee, M. Choi, T.-T. Kim, S. Lee, M. Liu, X. Yin, H. K. Choi, S. S. Lee, C.-G. Choi, S.-Y. Choi, X. Zhang, and B. Min, "Switching terahertz waves with gate-controlled active graphene metamaterials," Nat. Mater. **11**, 936–941 (2012).